\documentclass[aps,physrev,twocolumn,nofootinbib, superscriptaddress,10pt,floatfix]{revtex4-2}
\usepackage{amsmath,amsthm,amsfonts,amssymb,amscd,bbold,mathrsfs,bm}
\usepackage{graphicx}
\usepackage{xcolor}
\usepackage{dcolumn}
\usepackage{physics}
\usepackage{hyperref}
\begin{document}
\title{Quantum error correction assisted axion search in CMOS spin qubit arrays}
\author{Xiangjun Tan}
\email{xiangjun.tan.25@ucl.ac.uk}
\affiliation{Department of Physics and Astronomy, University College London, WC1E 6BT London, United Kingdom}
\author{Zhanning Wang}
\affiliation{Instituto de Ciencia de Materiales de Madrid (ICMM), Consejo Superior de Investigaciones Cient\'ificas (CSIC), 28049 Madrid, Spain}
\date{\today}

\begin{abstract}
Searches for axion and axionlike dark matter based on solid-state spin qubits are fundamentally limited by strong longitudinal dephasing, which rapidly suppresses the sensitivity gains offered by entanglement.
Here we show that quantum error correction (QEC) can substantially enhance axion search sensitivity in realistic semiconductor spin qubit platforms by mitigating this dominant noise source.
By integrating an optimally chosen repetition code QEC with logical GHZ block entanglement, we derive closed-form expressions for the quantum Fisher information that explicitly incorporate the finite coherence time of the axion field.
Our analysis demonstrates that modest QEC cycle frequencies are sufficient to significantly reduce the effective dephasing rate, thereby restoring a broad parameter regime in which entanglement-enhanced sensing surpasses the standard quantum limit.
Projecting these results onto CMOS-compatible device parameters, we find that QEC-protected entangled sensing can revive otherwise inaccessible quantum advantages, yielding up to order-of-magnitude improvements in sensitivity to the axion-electron coupling $g_{ae}$.
These results establish a practical and theoretically controlled pathway for using QEC to improve qubit array searches for physics beyond the Standard Model.
\end{abstract}
\maketitle

\section{Introduction}
\label{Sec: Introduction}

The microscopic nature of dark matter (DM) remains a central topic in astrophysics \cite{Feng2010, Marsh2016, Luca2020, Choi2021, Arbey2021}.
One of the most studied candidates is the QCD axion, which was introduced to solve the strong CP problem of quantum chromodynamics \cite{Peccei1977, Peccei1977_2, Weinberg1978, Wilczek1978, Berezhiani1991, Khlopov1999}.
In this picture the CP-violating vacuum angle $\bar{\theta}$ is promoted to a dynamical field $a(x)$ associated with a spontaneously broken Peccei-Quinn symmetry \cite{Kim2010_RMP, Irastorza2018}.
Non-perturbative QCD effects generate a periodic potential that drives the axion field to a CP-conserving minimum and explains the smallness of the neutron electric dipole moment \cite{Duffy2009, Luca2020}.
If the axion is produced by the vacuum misalignment mechanism, its coherent oscillations behave today as cold DM \cite{Preskill1983, Abbott1983}.
In the galactic halo the axion is well described as a classical, spatially coherent field with very large occupation number, with a central oscillation frequency set by the axion mass $m_a$ and a coherence time of order $(m_a v^2)^{-1}$, where $v\sim 10^{-3}c$ is the virial velocity \cite{Dine1983, Francesca2022, Brady2022, Guo2024}.

Besides its coupling to photons, the axion can couple to electrons.
At low energies this interaction appears as a derivative coupling to the axial current \cite{Stadnik2014, Yannis2022}.
In the non-relativistic limit relevant for galactic DM, the spatial gradient of the axion field acts as an effective oscillating pseudo-magnetic field that couples to the electron spin \cite{Balakin2015, Berlin2024}.
This axion wind can drive spin precession at a frequency set by $m_a$, typically in the radio or microwave range in QCD axion models \cite{Garcon2018, Garcon2019}.
Detecting this effect requires phase coherent control and readout over times that are long compared to typical solid state coherence times but still shorter than the axion coherence time \cite{Dror2023, Zitong2024, Xiangjun2026_APL}.

Several existing and proposed experiments probe axion-spin couplings with macroscopic spin systems, but the relevant signal and noise mechanisms differ between platforms.
Nuclear-spin NMR searches, including CASPEr and related liquid- and solid-state implementations, measure the precession of a polarized spin ensemble driven by the axion-induced effective field \cite{Teng2019, Budker2014, Aybas2021}.
In such NMR-like experiments, the usable signal is set by the available spin polarization or sample magnetization, while transverse coherence, inhomogeneous broadening, spin-projection noise and readout noise determine the product-state SQL benchmark \cite{Aybas2021_QST, Blanchard2023, Boyers2025}.

Ferromagnetic haloscopes constitute a distinct class \cite{Barbieri2017, Crescini2018, Crescini2020}.
Rather than measuring the precession of an ensemble of independent spins, these experiments search for the resonant excitation of a collective magnetic mode, such as the uniform Kittel magnon mode, driven by the axion-induced effective field \cite{Chigusa2020, Chigusa2024}.
The magnon signal is transduced through a microwave cavity or photon-magnon hybrid system and measured with near quantum-limited microwave or heterodyne readout \cite{Flower2019, Clerk2010}.
The relevant quantum noise is therefore associated with phase-insensitive resonant detection and the vacuum or residual thermal occupation of the magnon-photon modes, not with the thermal magnetization and spin-projection SQL of uncorrelated spins \cite{Ikeda2022, Clarence2026}.
In the remainder of this work, the term SQL refers to the product-state benchmark for the semiconductor spin qubit array considered here, using the same number of physical spins as the QEC-protected entangled protocol.

Recent progress in quantum information science suggests a different approach, in which engineered qubits or solid state quantum devices are used directly as DM sensors \cite{Dixit2021, Chen2023, Andrew2023, Saebyeok2024, Chen2024, Saebyeok2025, Sungjae2025, Braggio2025}.
Gate defined semiconductor spin qubits are a natural candidate for this role \cite{Zwanenburg2013_RMP, Burkard2023_RMP, Guangchong2025}.
Isotopically enriched Si/SiGe and related heterostructure support long coherence times and high fidelity one- and two-qubit gates \cite{Zhao2019, Noiri2022, Xue2022}.
The devices are compatible with complementary metal-oxide-semiconductor (CMOS) technology and can be arranged in dense arrays \cite{Zwerver2022, Lawrie2023, Xander2025, Steinacker2025}.
In systems with strong spin-orbit coupling, the effective $g$-factor can be tuned electrically, which allows rapid, on chip tuning of the qubit resonance frequency to follow an unknown axion mass \cite{Liles2021, ik2024, Xiangjun2026}.
In current semiconductor devices, where noise is dominated by local fluctuations in the longitudinal magnetic field, large GHZ states quickly lose coherence and do not outperform an SQL-limited protocol \cite{Huelga1997, Demkowicz2012, Yoneda2018}.

Quantum error correction (QEC) offers a way to protect entangled states from noise, but its use in sensing is subtle \cite{Giovannetti2006, Kessler2014, Arrad2014, Degen2017_RMP}.
Semiconductor spin qubits coupled to an axion-like field provide a concrete example of this situation, where we search for the in-plane qubit spin polarizations, while the noise is mainly longitudinal \cite{Hajime2025, Shion2025}.
The repetition code suppresses stochastic local phase errors when those errors are syndrome-visible, while the QEC-enhanced sensing channel is formulated as a transverse or dressed-frame code-preserving logical generator.
The de Broglie wavelength of the non-relativistic axion field is of order $10^2$-$10^3$~m for typical parameters, so it acts as a nearly uniform drive across a nanometer scale qubit array, while charge noise, phonons, and residual hyperfine interactions are local and only weakly correlated between distinct qubits \cite{Centers2021, Brady2022, Wenming2024}.

In this work we consider a simple and device compatible protocol based on a repetition code that corrects local phase flip ($Z$) errors but leaves a global transverse $\sigma_x$ signal unaffected.
We encode each logical qubit in $n_{\text{rep}}$ physical spins and form logical GHZ states of size $k_L$ from these logical qubits.
The total number of physical spins in a sensing block is then $N_{\text{phys}} = n_{\text{rep}} k_L$.
During the axion-sensitive free-evolution period, repeated QEC cycles suppress the effect of uncorrelated dephasing, which is described by a physical rate $\gamma_{\text{loc}}$, and replace it with an effective logical dephasing rate $\gamma_{\text{eff}}(n_{\text{rep}})$.
This rate reflects both the binomial suppression of physical phase flips by the code and additional errors from imperfect gates and measurements.
Correlated noise enters as a separate rate $\gamma_{\text{cor}}$ that is common to all qubits and is not corrected by the repetition code.
We also include the finite coherence time of the axion field, so that the interrogation time in each sensing segment is set by the minimum of the hardware limit and $\tau_a(m_a)$.

Within this framework we derive closed expressions for the quantum Fisher information of QEC-protected GHZ blocks and for the corresponding gain over an SQL limited protocol with the same number of physical spins.
The resulting formulas show explicitly how the sensitivity depends on $N_{\text{phys}}$, on $\gamma_{\text{cor}}$ and $\gamma_{\text{eff}}(n_{\text{rep}})$, and on the axion limited interrogation time.
For each axion mass we identify an optimal logical GHZ size $k_L^*(m_a)$ that balances signal amplification against dephasing of the entangled state.
Using representative parameters for CMOS compatible semiconductor spin qubits, we find that modest QEC cycle frequencies, which reduce the effective uncorrelated dephasing rate by about one order of magnitude, are enough to restore a regime where multi-qubit entanglement improves sensitivity beyond the SQL.
In this regime the protocol achieves an approximately constant improvement factor in the projected sensitivity to the axion–electron coupling $g_{ae}$ across a broad mass range, while using the same physical hardware as a conventional sensor.

Although we focus on gate defined semiconductor spin qubits, the basic idea is more general.
Any qubit platform in which a code-preserving collective signal can be distinguished from syndrome-visible local dephasing noise can implement a similar selective QEC strategy.
Our results therefore suggest a practical way to integrate quantum error correction into qubit-based searches for axion DM and other forms of physics beyond the Standard Model.

\begin{figure}[htbp!]
\centering
\includegraphics[width=\columnwidth]{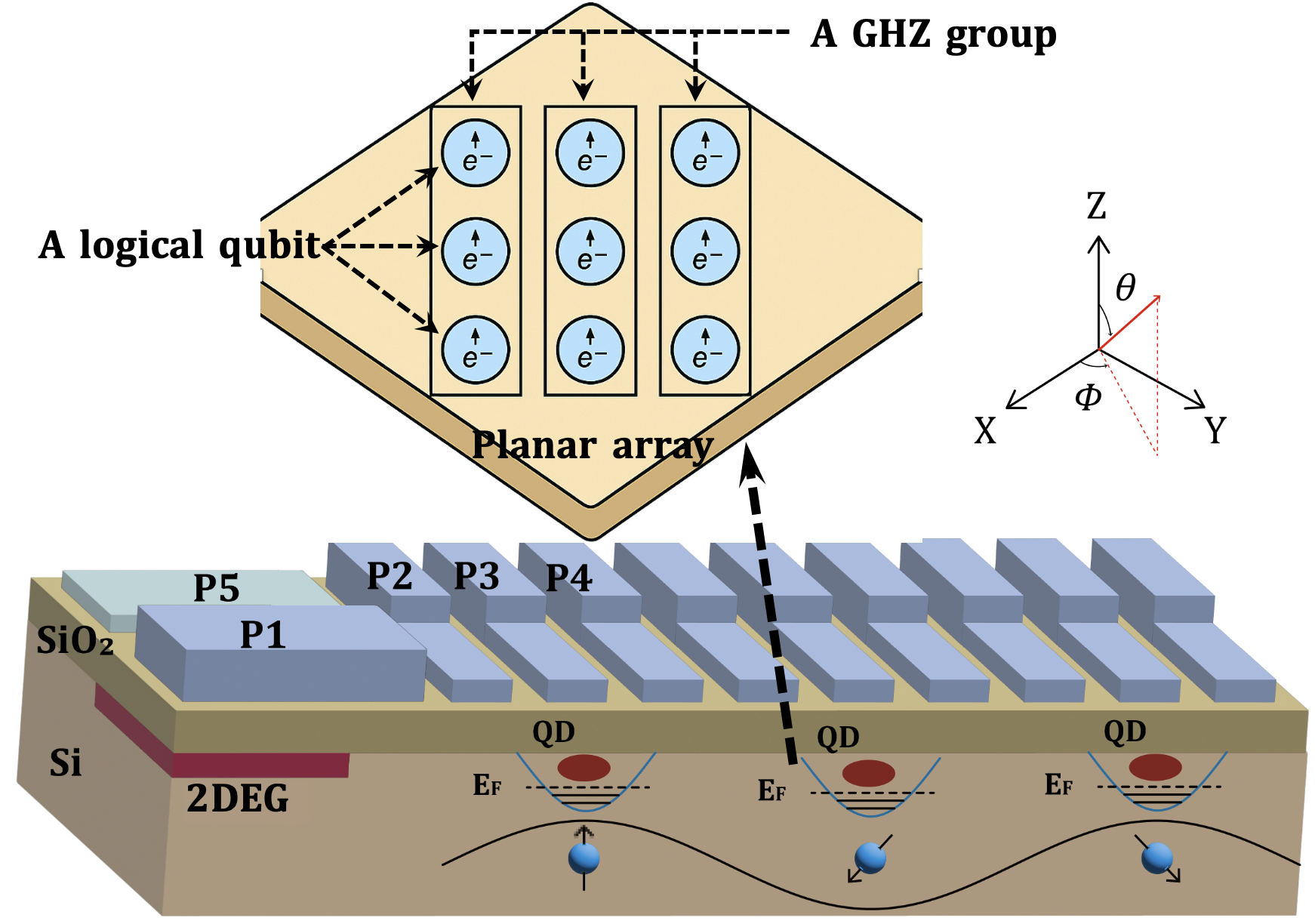}
\caption{
A schematic planar silicon quantum dot array.
In this specific design, we focus on a quantum dot array in the silicon layer.
By applying a gate electric field via gate P1, electrons accumulate in silicon and are confined vertically against the silicon oxide (indicated at the location of the two-dimensional electron gas).
The single quantum dot is formed using gates P2-P5.
The gates P2 and P4 provide confinement in the x-direction, while gate P5 provides confinement in the y-direction.
P3 is used as the top gate of the quantum dot, accumulating a single electron in the potential well beneath.
Then the quantum dot array repeats the set-up for P2-P4.
In the upper half of the figure, we sketch the planar qubit array, where each blue circle represents a physical electron spin qubit, and vertical columns denote repetition code blocks implementing a logical qubit.}
\label{fig:Array}
\end{figure}
\section{Model and methodology}
\label{Sec: Model and methodology}
We start our exploration of the axion-electron coupling with the Lagrangian density that describes the interaction between axions and electrons.
Through our analysis, we adopt the natural units $\hbar = c = 1$ for simplicity, and the Minkowski metric with the signature $(+,-,-,-)$.
The axion field is represented by the pseudoscalar field $a(x)$, while the electron field is denoted by the Dirac spinor field $\psi(x)$.
The interaction Lagrangian density for the axion-electron coupling can be expressed as:
\begin{equation}
\mathcal{L}_{ae} = \frac{C_e}{2 f_a} \partial_\mu a \, \bar{\psi} \gamma^\mu \gamma^5 \psi \,,
\end{equation}
where $C_e$ is a dimensionless coupling constant that characterizes the strength of the interaction, and $f_a$ is the axion decay constant with dimensions of energy.
The conjugation of the Dirac spinor field is given by $\bar{\psi} = \psi^\dagger \gamma^0$, and $\gamma^\mu$ are the gamma matrices satisfying the Clifford algebra, $\{\gamma^\mu, \gamma^\nu\} = 2 g^{\mu\nu}$, with $\gamma^5 = \mathrm{i}\,\gamma^0 \gamma^1 \gamma^2 \gamma^3$.
For simplicity, we define the coupling constant $g_{ae} \equiv C_e / (2 f_a)$, which has dimensions of inverse energy.
A dimensionless quantity can be defined as $\tilde{g}_{ae} \equiv 2 g_{ae} m_e$, where $m_e$ is the electron mass, to facilitate comparisons with experimental constraints.

We focus on the mass range of the axion that is mostly relevant for cold light DM candidates, typically in the range of $10^{-6}$~eV to $10^{-3}$~eV \cite{Yannis2022}.
In this mass range, the axion field can be treated as a classical field oscillating at a frequency determined by its mass.
The axion field satisfies the Klein-Gordon equation, and for a non-relativistic axion field, we can express it as:
\begin{equation}
a(t,\bm{x}) = a_0 \cos(\omega_a t - \bm{k}_a \cdot \bm{x} + \phi) \,,
\end{equation}
where $a_0$ is the amplitude of the axion field, $\omega_a$ is the axion angular frequency, $\bm{k}_a$ is the axion wave vector, and $\phi$ is a random phase.
The velocity of the axion DM in the galactic halo is typically $v_a \sim 10^{-3}$, leading to a small momentum $\bm{k}_a = m_a \bm{v}_a$.
The axion angular frequency is related to its mass $m_a$ by:
\begin{equation}
\omega_a = m_a + \frac{\bm{k}_a^2}{2 m_a} \approx m_a \,.
\end{equation}
Assuming the axion constitutes the entirety of the local DM density $\rho_{\text{DM}} \approx 0.3-0.45$~GeV/cm$^3$, the amplitude $a_0$ can be related to $\rho_{\text{DM}}$ by:
\begin{equation}
\rho_{\text{DM}} = \frac{1}{2} m_a^2 a_0^2 \Rightarrow a_0 = \sqrt{\frac{2 \rho_{\text{DM}}}{m_a^2}} \,.
\end{equation}

Axion frequency broadening from the axion DM velocity distribution is also important for experimental searches in qubit architectures, whose dephasing times can range from nanoseconds to milliseconds \cite{Gramolin2022, Shion2024}.
The velocity distribution of the axion DM in the galactic halo can be approximated by a Maxwell-Boltzmann distribution with a characteristic velocity $v_0 \sim 220$~km/s \cite{Aaron2022}.
The frequency broadening $\Delta \omega_a$ follows $\Delta \omega_a / \omega_a \sim v_0^2 \sim 10^{-6}$, leading to the following coherence parameters for the axion field:
\begin{equation}
\tau_a \sim \frac{1}{m_a v_0^2}, \quad \ell_a \sim \frac{1}{m_a v_0} \,,
\end{equation}
where $\tau_a$ is the axion coherence time and $\ell_a$ is the axion coherence length.
For an axion mass of $m_a \sim 10^{-6}$~eV, the coherence time is $\tau_a \sim 1$~ms, and the coherence length is $\ell_a \sim 300$~m.
As an estimation, if the qubit working time (dephasing time) $T_2^*$ is much shorter than the axion coherence time $\tau_a$, the axion field can be treated as a coherent monochromatic oscillation during the qubit operation.
Meanwhile, since the qubit array size is much smaller than the axion coherence length $\ell_a$, the spatial variation of the axion field across the qubit array can be neglected.

For a broader axion mass range, where the axion coherence time is comparable to the qubit dephasing time, the sensing strategy needs to be adjusted accordingly to account for the partial coherence of the axion field during the qubit operation, which is further discussed in the following sections.

Next, we need to consider the low-energy effective Hamiltonian describing the electron and axion system, in preparation for the analysis of the axion signals.
Once we have solved for the axion field $a(t,\bm{x})$, we can perform a variation with respect to $\bar{\psi}$ of the Lagrangian density $\mathcal{L}_e + \mathcal{L}_{ae}$:
\begin{equation}
\bar{\psi}(\mathrm{i}\,\gamma^\mu\partial_\mu - m_e) \psi + g_{ae} (\partial_\mu a) \bar{\psi} \gamma^\mu \gamma^5 \psi \,.
\end{equation}
Then we can derive the following Hamiltonian:
\begin{equation}
H = \bm{\alpha} \cdot \bm{p} + \beta m_e - g_{ae} (\partial_\mu a) \gamma^0 \gamma^\mu \gamma^5 \,,
\end{equation}
where $\alpha^i = \gamma^0 \gamma^i$, and $\beta = \gamma^0$.
Consider the four component nature of the Dirac spin field $\psi=[\varphi_1, \varphi_2]^T$, we can separate the non-relativistic component $\varphi$ and small part $\varphi_2$.
A Schrieffer-Wolff transformation to the total Hamiltonian $H$ will lead to the low-energy non-relativistic Hamiltonian:
\begin{equation}
H = \frac{\bm{p}^2}{2m_e} - \frac{g_{ae}}{m_e} (\partial_t a) \bm{\sigma} \cdot \bm{p} + \frac{g_{ae}^2 (\partial_t a)^2}{2m_e} - g_{ae} \nabla a \cdot \bm{\sigma} \,.
\end{equation}
The first term is the kinetic energy term of the electron, the second term is the spin-momentum coupling, the third term is a spin irrelevant diagonal energy shift, and the last term is the spin-axion gradient coupling.
A simple comparison between the second and the fourth term will yield a ratio $|\bm{p}|/(m_e v)$.
In our case of a localized electron in a quantum dot, the magnitude of the momentum will be small, indicating that the spin-gradient term will dominate.
Therefore, we obtain the working Hamiltonian for searching for the light cold dark axion in the spin qubit platform:
\begin{equation}
H_{\text{eff}} = \frac{\bm{p}^2}{2m_e} - g_{ae} \nabla a \cdot \bm{\sigma} \,,
\end{equation}
where the second term is referred to as the axion wind term.
This Hamiltonian spans the major components of searching the axion signals using semiconductor quantum dot qubit platform, with possible extensions to the electron spin qubit, hole spin qubit, and the nuclear spin qubit.

\subsection{Axion-electron Spin Qubit Hamiltonian}
\label{Subsec: Axion-qubit Hamiltonian}

We consider an electron spin qubit confined in a semiconductor
heterostructure or CMOS device, where the orbital degree of freedom is
frozen into an $s$-like ground state defined by the gate potential.
An external static magnetic field $B_0\hat{\mathbf{z}}$ sets the
quantization axis, giving the Zeeman Hamiltonian $H_0=\omega_0\sigma_z/2$ with $\omega_0 = g_e \mu_B B_0$, where $g_e=2$.

Now we can evaluate the gradient of the axion field, from which the amplitude and the direction of the axion wind $\hat{\text{n}}$ can be extracted as:
\begin{equation}
|\nabla a| = m_a v a_0 = v \sqrt{2\rho_{\text{DM}}}\,, \quad \hat{\mathbf{n}} = \frac{\nabla a}{|\nabla a|} \,.
\end{equation}
We can define the relative angle between the axion wind and the qubit quantization axis as $\cos \theta = \hat{\mathbf{n}} \cdot \hat{\mathbf{z}}$.
We keep the orientation of the axion wind relative to the qubit quantization axis explicit.
Writing
\begin{equation}\label{Eq:axion_wind_orientation}
\hat{\mathbf{n}} = \cos\theta\,\hat{\mathbf{z}} + \sin\theta\left(\cos\varphi_\perp\,\hat{\mathbf{x}}+\sin\varphi_\perp\,\hat{\mathbf{y}}\right),
\end{equation}
the axion-induced spin Hamiltonian can be decomposed as
\begin{equation}\label{Eq:axion_decomposition}
H_a(t)=\frac{\Omega_a(t)}{2} \left[ \cos\theta\,\sigma_z + \sin\theta\left(\cos\varphi_\perp\,\sigma_x+\sin\varphi_\perp\,\sigma_y\right) \right],
\end{equation}
with
\begin{equation}\label{Eq:Omegaa_general}
\Omega_a(t)=-2g_{ae}v\sqrt{2\rho_{\text{DM}}}\cos(\omega_a t+\phi).
\end{equation}
The longitudinal component proportional to $\sigma_z$ produces the phase-modulation sideband response analyzed in Refs.~\cite{Tan2025IEEE,Tan2025PRD}.
 For the sideband formulas immediately below we set $|\cos\theta|=1$ for compactness; a general orientation only multiplies the longitudinal susceptibility by $|\cos\theta|$.
 The QEC-protected analysis introduced later instead uses the transverse component, or an equivalent dressed-frame implementation, because that channel can be made compatible with the phase-flip repetition code.

For simplicity, we consider the case where we select the external magnetic field $\bm{B}$ to be parallel to the axion wind direction $\hat{\mathbf{n}}$, therefore, the total Hamiltonian will read:
\begin{equation}
H_{\text{total}} = \frac{\omega_0}{2}\sigma_z - g_{ae} v \sqrt{2\rho_{\text{DM}}} \cos(\omega_at + \phi) \sigma_z \,,
\end{equation}
where we have absorbed the phase shift and $\varphi$ into the new phase factor $\phi$.
From this Hamiltonian, we can define the effective magnetic field induced by the axion oscillations:
\begin{equation}
B_{\text{eff}} = -\frac{2g_{ae}}{g_e \mu_B} v \sqrt{2\rho_{\text{DM}}} \cos(\omega_a t + \phi) \,.
\end{equation}
A typical estimation of the magnitude of this effective magnetic field is:
\begin{equation}
4 \times 10^{-22}~\text{T} \left(\frac{g_{ae}~\text{GeV}}{10^{-11}}\right) \left(\sqrt{\frac{\rho_{\text{DM}}~\text{cm}^3}{0.4~\text{GeV}}}\right) \left(\frac{v}{10^{-3}}\right) \,.
\end{equation}
This extremely small effective magnetic field motivates the use of quantum-enhanced sensing techniques to improve the sensitivity of axion searches.
Most state-of-the-art search platforms rely on the following advantages, including large number of spins, long coherence time, and low-noise readout techniques, where quantum error correction can, in principle, extend coherence.

The transition from the astroparticle axion signal to the qubit response starts from this time varying effective magnetic field.
The searching strategy we consider centers around the phase variation of the spin polarization.
The spin populations will not change due to the commutation of $\sigma_z$ with the total Hamiltonian, therefore, we need to initialize the qubit in the transverse plane, e.g., the $\sigma_x$ eigenstate $\ket{\psi;t=0} = (\ket{\uparrow} + \ket{\downarrow})/\sqrt{2}$.
Under the influence of the axion effective magnetic field, the qubit state will pick up a total time-dependent phase:
\begin{equation}
\phi_a(t) = \omega_0 t - \beta \sin(\omega_a t) \,,
\end{equation}
where we have selected the initial phase $\phi = 0$ for simplicity, and the modulation index $\beta$ is defined as:
\begin{equation}
\beta \equiv \frac{2 g_{ae} v \sqrt{2\rho_{\text{DM}}}}{m_a} \,.
\end{equation}
This is a standard frequency modulation form, with the carrier frequency $\omega_0$, and modulation frequency $\omega_a$.
If we evaluate $\sigma_x$ expectation value, we will obtain $\expval**{\sigma_x(t)} = \cos(\omega_0 t - \beta \sin(\omega_a t))$.
Using the Jacobi-Anger expansion, we can rewrite this expectation value as:
\begin{equation}
\expval**{\sigma_x(t)} = \sum_{n=-\infty}^{\infty} J_n(\beta) \cos[(\omega_0-n\omega_a)t] \,.
\end{equation}
This expression shows that the axion-electron coupling induces sidebands at frequencies $\omega_0 \pm n \omega_a$ with amplitudes determined by the Bessel functions $J_n(\beta)$ as shown in Ref.~\cite{Tan2025IEEE}.

Now, we need to integrate this phase accumulation with the experimental considerations.
At the device level, the qubit dephasing time $T_2^*$ should be considered, which limits the maximum free evolution time.
This is due to the environmental noise, including charge noise, nuclear hyperfine interactions in the semiconductor host materials, which can be re-focused using dynamical decoupling and spin echo techniques.
In our analysis, these effects are captured phenomenologically by a $1/f$ spectrum with power spectral density $S(\omega) \propto 1/\omega^\alpha$, where $\alpha \sim 0.7-1$ for semiconductor spin qubits, and the device dephasing time $T_2^*$ is selected from Ref.~\cite{Wang2025}.

Assuming the whole experiment runs for a total time $T_{\text{tot}}$, we can divide the total time into $N_{\text{seg}}$ segments, each with a free evolution time of $T_{\text{seg}}$, which is limited by both the hardware and the axion coherence time $\tau_a$:
\begin{equation}
T_{\text{seg}}(m_a) = \min\{ T_2^*, \tau_a(m_a) \} \,.
\end{equation}
The number of segments is then $N_{\text{seg}} = T_{\text{tot}} / T_{\text{seg}}(m_a)$, and each segment produces an independent measurement of the axion-induced phase accumulation, known as the semi-coherent search.
To connect with standard magnetometer benchmarks, we note that for a single-qubit magnetometer operating at the standard quantum limit, the signal-to-noise ratio for detecting a static effective field $B_{\text{eff}}$ after a total integration time $t$ is given by:
\begin{equation}
\text{SNR}_{\text{SQL}}(t) = \frac{B_{\text{eff}} T_{\text{seg}}}{\eta_B \sqrt{T_{\text{seg}}}} \sqrt{\frac{t}{T_{\text{seg}}}} = \frac{B_{\text{eff}} \sqrt{t}}{\eta_B} \,,
\end{equation}
where $\eta_B$ is the field sensitivity in units of T/$\sqrt{\text{Hz}}$.
It includes the conversion of a magnetic perturbation into a spin phase, spin-to-charge readout, resonator or amplifier noise, and the averaging protocol.
Equivalently, a field $B_{\text{eff}}$ gives an SQL signal-to-noise ratio $B_{\text{eff}}\sqrt{t}/\eta_B$ after total integration time $t$.
Changing $\eta_B$ rescales the absolute projected coupling sensitivity, but it does not change the dimensionless QEC gain obtained from the Fisher-information ratio below.
The representative value used in Tab~\ref{Tab: device_params} should therefore be interpreted as a hardware-level benchmark for semiconductor spin qubit magnetometry \cite{Tan2025IEEE,Tan2025PRD}.

\subsection{Dephasing channels}
\label{Subsec: Dephasing channels}
Now we focus on the dephasing channels that affect the qubit coherence during the free evolution time, in preparation for the discussion of the quantum error correction protocol.
In semiconductor quantum dot devices, noise is represented as a diagonal energy drift term $\delta\omega_N(t)$, which leads to the single total Hamiltonian:
\begin{equation}
H_{\text{total}} = \frac{\omega_0}{2}\sigma_z + \frac{\Omega_a(t)}{2} \sigma_z + \frac{\delta\omega_N(t)}{2} \sigma_z \,,
\end{equation}
where $\Omega_a(t) = -2 g_{ae} v \sqrt{2\rho_{\text{DM}}} \cos(\omega_a t + \phi)$, and $\delta\omega_N(t)$ is the longitudinal noise term (Z error), which leads to an effective dephasing envelope characterized by a timescale $T_2^*$ determined by the statistics of $\delta\omega_N(t)$.

The microscopic evaluation of this noise term must consider the device details \cite{Abadillo2023}, while we focus on a phenomenological description here:
\begin{equation}
\delta\omega_N(t) = \delta\omega_{\text{cor}}(t) + \delta\omega_{\text{loc}}(t) \,,
\end{equation}
where $\delta\omega_{\text{cor}}(t)$ is the correlated noise term that affects all qubits in the same way, such as due to the Overhauser field, while $\delta\omega_{\text{loc}}(t)$ is the local noise term that affects each qubit independently, such as random telegraph noise \cite{Reilly2008, Shalak2023, Megan2024}.
For an arbitrary noise realization, the spin polarization will pick up an dephasing envelope:
\begin{equation}
\overline{\expval**{\sigma_x(t)}} = \expval**{\mathrm{e}^{-\mathrm{i}\int_0^t \delta\omega_N(t') dt'}} \, \cos(\omega_0 t - \beta \sin(\omega_a t)) \,.
\end{equation}
Assuming the noise follows the Markovian approximation, the envelope factor can be expressed as $C(t)=\mathrm{e}^{-\gamma t}$, where $\gamma$ is the dephasing rate \cite{Degen2017_RMP}.
This form corresponds to the short-correlation-time limit of more general dephasing processes.

During a sensing segment of duration $T_{\text{seg}}$, experimentally relevant low-frequency noise can be captured by a stretched-exponential envelope
$C_{\text{loc}}(t)=\exp[-t/T_{2,\text{M}}-(t/T_{2,1/f})^{\beta}]$, which reduces to the Markovian form in the limit $T_2 \rightarrow \infty$ for the $1/f$ noise.

We emphasize that the repetition code suppresses stochastic local phase errors only when they are syndrome-visible.
The longitudinal Hamiltonian above is used to define the unencoded sideband response and the physical dephasing envelope.
In the encoded analysis below, the same local longitudinal device noise remains the syndrome-visible $Z_j$ channel, while the coherent axion response is assumed to be implemented as a transverse or dressed-frame code-preserving logical generator, as made explicit in Eqs.~\eqref{Eq:signal_code_space} and \eqref{Eq:logical_signal_ansatz}.
Although the correlated noise $\gamma_{\text{cor}}$ needs further treatment like decoherence-free subspace, it is typically much smaller than the local noise in semiconductor spin qubit devices thanks to isotopic enrichment, and the contact hyperfine interaction is even absent for hole spin qubits \cite{Itoh2014}.

For a concrete numerical analysis, we summarize the typical device parameters and operating conditions for Si/SiGe CMOS devices in Tab.~\ref{Tab: device_params}.
\begin{table}[htb]
\begin{ruledtabular}
\begin{tabular}{lc}
\textbf{Parameter} & \textbf{Value} \\
\hline
Magnetic sensitivity $\eta_B$ 
    & $100~\text{nT}/\sqrt{\text{Hz}}$ \\
Uncorrelated dephasing (Markov) $T_{2,\text{M}}$ 
    & $2~\text{ms}$ \\
Uncorrelated dephasing (1/f-like) $T_{2,1/f}$ 
    & $200~\mu\text{s}$ \\
Stretch exponent (1/f-like) $\beta$ 
    & $2$ \\
Correlated dephasing time $T_{2,\text{corr}}$     
    & $10~\text{ms}$ \\
QEC cycle time $\tau_{\text{cyc}}$                
    & $2~\mu\text{s}$ \\
Physical $Z$-error probability per cycle $p_Z$              
    & $\bigl[1-C_{\text{loc}}(\tau_{\text{cyc}})\bigr]/2$ \\
Gate error (future) $p_g$                                  
    & $10^{-4}\,(10^{-5})$ \\
Measurement error (future) $p_m$                           
    & $10^{-4}\,(10^{-5})$ \\
Logical dephasing rate $\gamma_{\text{eff}}$      
    & $2p_L(n_{\text{rep}})/\tau_{\text{cyc}}$ \\
Segment duration $T_{\text{seg}}$                 
    & $100~\mu\text{s}$ \\
Total physical qubits $N_{\text{phys}}$           
    & $n_{\text{rep}} k_L$ \\
\end{tabular}
\end{ruledtabular}
\caption{Device parameters and operating conditions used for the representative semiconductor electron spin qubit sensor.
The uncorrelated dephasing envelope is modeled as $C_{\text{loc}}(t)=\exp[-t/T_{2,\text{M}}-(t/T_{2,1/f})^{\beta}]$.
The millisecond-scale Markovian/Hahn-echo component $T_{2,\text{M}}=2~{\text{ms}}$ is motivated by recent silicon spin-qubit experiments with millisecond-scale coherence times \cite{Steinacker2025}.
The low-frequency component and stretch exponent are consistent with charge noise-limited semiconductor spin-qubit coherence \cite{Yoneda2018, Wang2025}.
The correlated dephasing time is the common-mode noise scale in isotopically enriched silicon
\cite{Itoh2014, Zhao2019}.
The control, cycle-time, and readout parameters are chosen to be compatible with recent high-fidelity exchange-based gates and fast PSB or charge-sensor readout in silicon multi-dot devices \cite{Lawrie2023, Steinacker2025}.
The magnetic sensitivity is an effective Ramsey magnetometry benchmark, estimated from the standard spin magnetometry scaling and including finite contrast, spin-to-charge conversion, readout infidelity and dead time \cite{Szechenyi2017}.}
\label{Tab: device_params}
\end{table}
Throughout this work we assume that the QEC cycle time $\tau_{cyc}$ is short compared to the characteristic correlation time of the dominant low-frequency noise, so that its effect can be coarse-grained into an effective per-cycle phase-flip probability.

\subsection{QEC with repetition code}
\label{Subsec: QEC and RC}

The faint frequency modulation signal is easily washed out by noise in the standard quantum limit, where a longer free evolution time can also accumulate phase errors due to noise, and the signal-to-noise ratio only scales with $\sqrt{N_{\text{shots}}}$, where $N_{\text{shots}}$ is the number of independent measurements.
The qubit array can be entangled into a GHZ state to enhance the signal accumulation, in the ideal decoherence-free limit, where the SNR scales with $N_{\text{phys}}$, the total number of physical qubits, and one must take quantum error correction into account to maintain the coherence of the entangled state.
In this subsection, starting from the device-level dephasing channels, we will first evaluate the logical error probability and corresponding dephasing rate.
This allows us to estimate how the GHZ coherence time is extended using quantum error correction, in preparation for the discussion of the Fisher information and axion coupling sensitivity in the following sections.

The QEC model below should be understood as a signal-preserving transverse or dressed-frame implementation.
A phase-flip repetition code corrects local $Z_j$ errors because these errors are syndrome-visible, but it does not preserve an arbitrary longitudinal signal.
For one repetition block,
\begin{equation}\label{Eq:signal_code_space}
\mathcal C= \text{span}\{ \ket{+}^{\otimes n_{\text{rep}}}, \ket{-}^{\otimes n_{\text{rep}}} \}, \qquad S_i=X_iX_{i+1}.
\end{equation}
If $P$ is the code-space projector, the relevant signal projections are
\begin{equation}\label{Eq:logical_signal_ansatz}
P\left(\sum_{j=1}^{n_{\text{rep}}}X_j\right)P = n_{\text{rep}}\bar Z, \qquad P\left(\sum_{j=1}^{n_{\text{rep}}}Z_j\right)P = 0 .
\end{equation}
Thus the encoded Fisher-information analysis applies to a collective transverse or pulse-level dressed-frame axion response, while the bare longitudinal sideband channel remains the unencoded baseline response.

We characterize the time cycle of an error correction round with $\tau_{\text{cyc}}$, which is taken to be $2~\mu$s in our analysis, including the syndrome measurement and recovery operations.
The probability of a physical $Z$ error occurring during each QEC cycle is given by:
\begin{equation}
p_Z = \frac{1}{2} \left[1 - C_{\text{loc}}(\tau_{\text{cyc}})\right] \,,
\end{equation}
which reduces to $p_Z = \gamma_{\text{loc}} \tau_{\text{cyc}}/2$ in the short cycle limit.
The number of physical qubits in each repetition code is denoted as $n_{\text{rep}}$, and the total number of physical qubits in the sensing block is $N_{\text{phys}} = n_{\text{rep}} k_L$, where $k_L$ is the number of logical qubits entangled into a GHZ state.
The optimal phase-flip error correction will correct up to $t = \lfloor (n_{\text{rep}}-1)/2 \rfloor$ physical $Z$ errors, where the $\lfloor f \rfloor$ defines the largest integer function of $f$.

To incorporate as much noise as possible, we also include the gate error $p_g$ and measurement error $p_m$ during each QEC cycle.
Up to the first order, the total logical $Z$ phase error probability $p_L(n_{\text{rep}})$ can be expressed as:
\begin{equation}
\sum_{k=t+1}^{n_{\text{rep}}} \binom{n_{\text{rep}}}{k} p_Z^k (1-p_Z)^{n_{\text{rep}}-k} + n_{\text{rep}} p_m + n_{\text{rep}} p_g \,.
\end{equation}
We expect that the logical error probability $p_L(n_{\text{rep}})$ can be significantly suppressed compared to the physical error probability $p_Z$ with an optimal choice of $n_{\text{rep}}$.
We notice that increasing $n_{\text{rep}}$ will also introduce complication of the QEC circuit and physics device overhead, and may also lead to higher correlated noise.

Now, we can extract the effective logical dephasing rate:
\begin{equation}\label{Eq:gammaeff}
\gamma_{\text{eff}}(n_{\text{rep}}) = -\frac{1}{\tau_{\text{cyc}}} \ln\left(1 - 2 p_L(n_{\text{rep}})\right) \approx \frac{2p_L(n_{\text{rep}})}{\tau_{\text{cyc}}} \,,
\end{equation}
which will replace the physical local dephasing rate $\gamma_{\text{loc}}$ during the free evolution time.
For a logical GHZ state of $k_L$ qubits, the envelope decay factor during a sensing segment of duration $T_{\text{seg}}$ reads:
\begin{equation}\label{Eq:C_overall}
\hspace{-10pt} C_{\text{GHZ}}(m_a) = \exp\left[-\left(\gamma_{\text{cor}} + \gamma_{\text{eff}}\right) k_LT_{\text{seg}}(m_a) \right] \,.
\end{equation}
Here $\chi$ is the single-spin susceptibility of the signal channel being compared.
In the QEC protocol it denotes the transverse or dressed-frame code-preserving axion response; in the unencoded sideband benchmark it denotes the longitudinal response.
Since both $F_{\text{GHZ}}$ and $F_{\text{SQL}}$ are proportional to $\chi^2$, the gain $\eta^2$ is independent of this normalization.
Performing the linear response of the axion coupling $g_{ae}$, we can define the susceptibility $\chi = \partial \phi_a / \partial g_{ae}$ at $g_{ae}=0$, which will lead to the following quantum Fisher information for the logical GHZ state:
\begin{equation}\label{Eq:FGHZ}
\mathcal{F}_{\text{GHZ}} = n_{\text{rep}}^2 k_L^2 \chi^2 \exp(-2(\gamma_{\text{eff}}+\gamma_{\text{cor}}) k_L T_{\text{seg}}) \,.
\end{equation}
For comparison, the SQL-limited Fisher information with the same number of physical qubits $N_{\text{phys}} = n_{\text{rep}} k_L$ reads:
\begin{equation}\label{Eq:FISQL}
\mathcal{F}_{\text{SQL}} = n_{\text{rep}} k_L \chi^2 \left[C_{\text{loc}}(T_{\text{seg}})\,\mathrm{e}^{-\gamma_{\text{cor}}T_{\text{seg}}}\right]^2 \,.
\end{equation}
From these two expressions, we define the Fisher-information gain over the SQL benchmark, $\eta^2 = F_{\text{GHZ}}/F_{\text{SQL}}$, as:
\begin{equation}
\eta^2 = n_{\text{rep}} k_L \frac{\mathrm{e}^{-2(\gamma_{\text{eff}} + \gamma_{\text{cor}}) k_L T_{\text{seg}}}}{C_{\text{loc}}^2(T_{\text{seg}}) \mathrm{e}^{-2 \gamma_{\text{cor}} T_{\text{seg}}}} \,.
\end{equation}
In the Markovian limit, $C_{\text{loc}}(T_{\text{seg}}) = \mathrm{e}^{-\gamma_{\text{loc}} T_{\text{seg}}}$, we have:
\begin{equation}
\eta^2 = n_{\text{rep}} k_L \mathrm{e}^{-2\left[(\gamma_{\text{eff}} + \gamma_{\text{cor}}) k_L - (\gamma_{\text{loc}} + \gamma_{\text{cor}})\right]T_{\text{seg}}} \,.
\end{equation}
This gain factor $\eta^2$ quantifies the improvement in sensitivity achieved by using QEC-protected GHZ states compared to the SQL limit with the same number of physical qubits.
Maximal gain can be achieved by optimizing the logical GHZ size $k_L$ for each axion mass $m_a$, balancing the signal amplification against the dephasing of the entangled state:
\begin{equation}\label{Eq:kstar_max}
k_L^* = \frac{1}{2(\gamma_{\text{eff}} + \gamma_{\text{cor}}) T_{\text{seg}}} \,.
\end{equation}
This closed-form optimum is obtained in the Markovian limit and serves as a useful guideline; under non-Markovian dephasing, the optimal block size is evaluated numerically using the same envelope formalism.
At this optimal point, the maximum Fisher information gain over the SQL limit is:
\begin{equation}
\eta_{\text{max}}^2 = n_{\text{rep}} k_L^* \frac{\exp[2(\gamma_{\text{loc}}+\gamma_{\text{cor}})T_{\text{seg}}]}{\mathrm{e}} \,.
\end{equation}
Guided by this optimal choice of $k_L^*$, we can evaluate the projected sensitivity to the axion-electron coupling $g_{ae}$ using QEC-protected GHZ states in the following sections.
\begin{equation}\label{Eq:Trep}
 T_{\text{rep}}^{(\alpha)}(m_a) = T_{\text{seg}}(m_a)+t_{\text{dead}}^{(\alpha)} \,,
\end{equation}
For a fixed wall-clock integration time $T_{\text{tot}}$, protocol $\alpha\in\{\text{GHZ},\text{SQL}\}$ can repeat only
\begin{equation}\label{Eq:Nseg}
N_{\text{seg}}^{(\alpha)}(m_a)=\frac{T_{\text{tot}}}{T_{\text{rep}}^{(\alpha)}(m_a)}
\end{equation}
times, with $T_{\text{rep}}^{(\alpha)}$ from Eq.~\eqref{Eq:Trep}.
The corresponding wall-clock Fisher information is therefore
\begin{equation}\label{Eq:Ftot}
\mathcal{F}_{\alpha}^{\text{(tot)}}(m_a)=\frac{T_{\text{tot}}}{T_{\text{rep}}^{(\alpha)}(m_a)}\,\mathcal{F}_{\alpha}(m_a),
\end{equation}
where $\mathcal{F}_{\alpha}$ denotes the segment-level quantity of Eqs.~\eqref{Eq:FGHZ} and \eqref{Eq:FISQL}.
The wall-clock gain becomes
\begin{equation}\label{Eq:etawall}
\eta_{\text{wall}}^2(m_a)=R_{\text{duty}}(m_a)\,\eta_{\text{seg}}^2(m_a),
\end{equation}
with
\begin{equation}\label{Eq:Rduty}
R_{\text{duty}}(m_a)\equiv \frac{T_{\text{rep}}^{\text{(SQL)}}(m_a)}{T_{\text{rep}}^{\text{(GHZ)}}(m_a)}.
\end{equation}
In the usual case $T_{\text{rep}}^{\text{(GHZ)}}\ge T_{\text{rep}}^{\text{(SQL)}}$, one has $R_{\text{duty}}\le 1$.
The plotted gain curves below focus on the matched-duty limit $R_{\text{duty}}=1$ in order to isolate the encoded-noise advantage.
Any extra inter-segment dead time of the encoded protocol then simply rescales the wall-clock payoff by $\sqrt{R_{\text{duty}}}$.
By contrast, we do not introduce an additional intra-cycle duty factor inside $T_{\text{seg}}$: if syndrome substeps interrupt signal accumulation within a correction cycle, the relevant modification is a filter-function problem rather than a constant prefactor, and that physics is absorbed here into the signal-preserving ansatz of Eq.~\eqref{Eq:logical_signal_ansatz}.
Equations~\eqref{Eq:gammaeff}-\eqref{Eq:Rduty} are the core outputs of the manuscript: once $\gamma_{\text{eff}}(n_{\text{rep}})$ and the repetition-period ratio are specified, they determine the encoded advantage as a function of device noise, control fidelity, axion coherence, and wall-clock overhead.
In the numerical curves below, however, $k_L$ is optimized over positive integers using the full envelope of Eq.~\eqref{Eq:C_overall}; Eq.~\eqref{Eq:kstar_max} is used only as a diagnostic guide.

\section{Results}
\label{sec:results}

In this section we quantify the impact of QEC-protected entangled sensing on the projected axion-electron coupling sensitivity, focusing on representative parameters for CMOS-compatible semiconductor spin qubits.
Unless stated otherwise, we take an uncorrelated dephasing time $T_2^{(\text{loc})}=2$~ms, a correlated dephasing time $T_2^{(\text{cor})}=10$~ms, and a hardware-limited interrogation time $T_{\text{lim}}=100$~$\mu$s.
The repetition distance $n_{\text{rep}}$ is varied explicitly when optimizing the QEC-assisted protocol.
For the representative projected sensitivity curves below, we use the optimized high-fidelity block $(n_{\text{rep}}, k_L) = (13, 14)$, corresponding to $\eta_{\text{seg}} \simeq 11.1$ in the matched-duty limit.
The local coherence and control/readout timescales are consistent with recent 
silicon spin qubit devices demonstrating millisecond-scale Hahn-echo coherence, 
high-fidelity exchange-based gates, and fast PSB or charge-sensor readout 
\cite{Noiri2022, Xue2022, Oakes2023, Steinacker2025}.
The correlated dephasing time is treated as a phenomenological baseline parameter.

\begin{figure}[htbp!]
\centering
\includegraphics[width=\columnwidth]{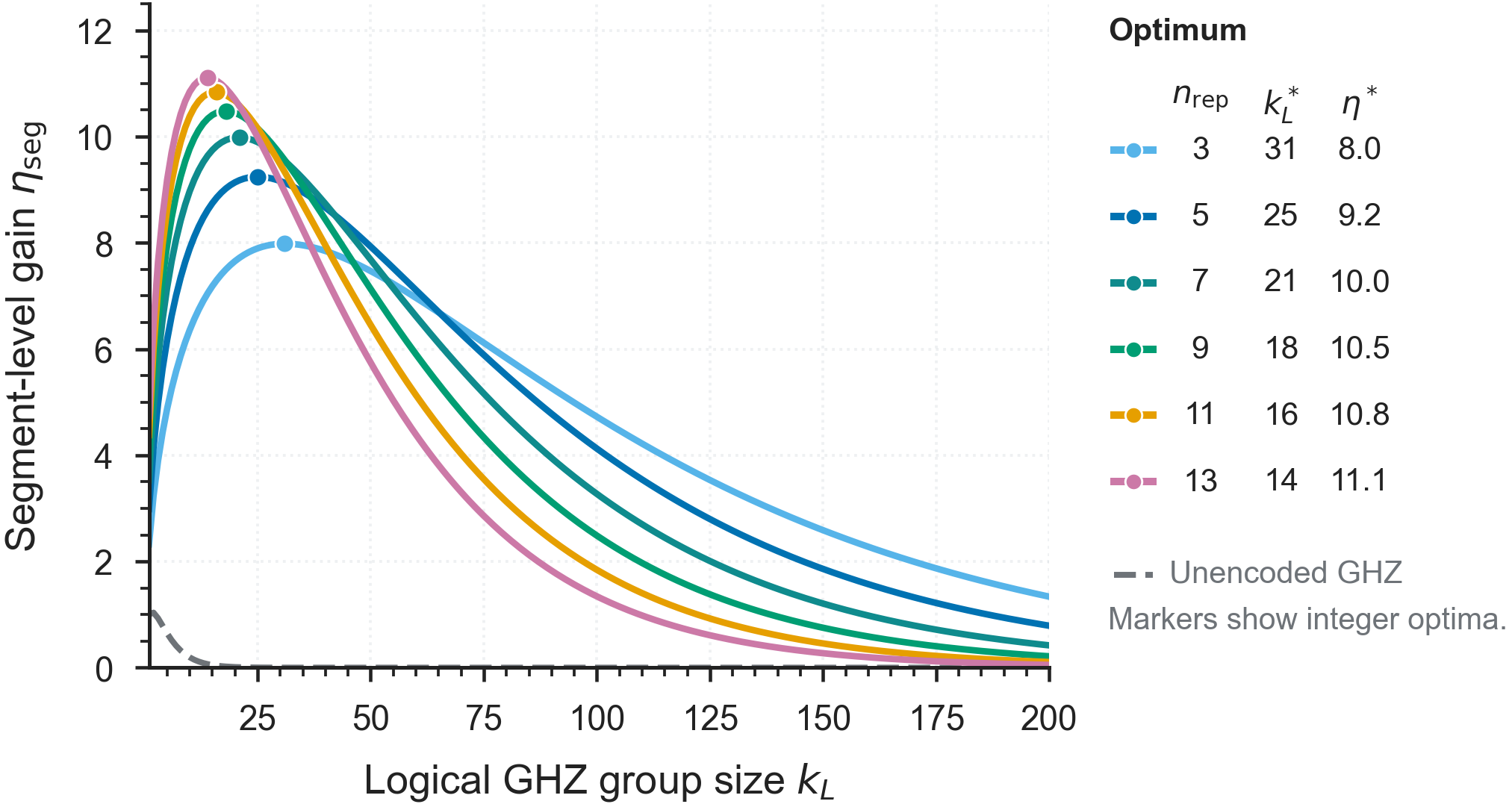}
\caption{Sensitivity improvement gain factor $\eta(k_L)$ for QEC-protected GHZ blocks of logical size $k$, evaluated for future possible semiconductor spin qubit dephasing parameters.
Curves correspond to repetition codes of distance $n_{\text{rep}}=3-13$, with peaks marked for the high confidence distances $n_{\text{rep}}=9-13$.
The maximal enhancement $\eta_{\max}$ grows monotonically with code distance, while the optimal GHZ size $k$ decreases, reflecting suppression of uncorrelated dephasing by QEC.
For comparison, the unencoded GHZ (grey) achieves only a weak enhancement and rapidly becomes worse than SQL for large $k$, whereas QEC recovers a finite and robust entangled sensing window.
Uncorrelated dephasing is modeled using a stretched-exponential envelope, with the Markovian limit recovered for $T_{2,1/f}\rightarrow\infty$.}
\label{fig:optimal_k}
\end{figure}
\subsection{Regimes of QEC benefit and limitations}
\label{subsec:results_regimes}

The analytic expressions derived in Sec. II allow us to identify the parameter regions in which QEC-assisted entangled sensing provides a meaningful advantage.
 For unencoded GHZ states, the relevant quantity is the dephasing accumulated per logical qubit, $(\gamma_{\text{loc}}+\gamma_{\text{cor}})T_{\text{seg}}(m_a)$.
Whenever $(\gamma_{\text{loc}}+\gamma_{\text{cor}}) T_{\text{seg}}(m_a) \gtrsim 0.1$, the optimal GHZ size inferred from Eq.~\eqref{Eq:kstar_max} is reduced to only a few logical qubits, and the practical entanglement advantage over SQL-limited sensing becomes strongly suppressed.
When the full dephasing envelope of Table~\ref{Tab: device_params} is used, including the $1/f$-like local component, this regime is reached for representative present-day semiconductor spin qubit parameters with millisecond-scale Hahn-echo coherence and $T_{\text{seg}}\sim100~\mu{\text{s}}$ \cite{Yoneda2018, Steinacker2025}.

Repetition-code QEC modifies this condition by replacing the single-qubit dephasing rate $\gamma_{\text{loc}}$ with the logical rate $\gamma_{\text{eff}}(n_{\text{rep}})$ obtained from Eq.~\eqref{Eq:gammaeff}.
The optimal block size becomes
\begin{equation}\label{Eq:kstar_restate}
k_L^*(m_a)=\frac{1}{2\big(\gamma_{\text{cor}}+\gamma_{\text{eff}}(n_{\text{rep}})\big)T_{\text{seg}}(m_a)}.
\end{equation}
Thus QEC produces a genuine metrological advantage whenever
\begin{equation}\label{Eq:QEC_condition}
(\gamma_{\text{cor}}+\gamma_{\text{eff}}(n_{\text{rep}}))T_{\text{seg}}(m_a)\lesssim 0.05 \,,
\end{equation}
which restores $k_L \gg 1$ and enables multi-qubit GHZ enhancement.
The effective logical dephasing rate $\gamma_{\text{eff}}(n_{\text{rep}})$ is set by a balance between the binomial suppression of physical dephasing and the additional gate and readout faults introduced by the code.
As a result, the attainable GHZ size is determined not by a single suppression parameter but by the combined decay scale $(\gamma_{\text{cor}}+\gamma_{\text{eff}})$, which fixes the optimal block length $k_L^*$.

Correlated dephasing affects all protocols by increasing the overall decay rate in Eq.~\eqref{Eq:C_overall}, but it does \textit{not} eliminate the relative QEC advantage: both encoded and unencoded coherence functions contain the same $\gamma_{\text{cor}}$, while QEC selectively reduces the logical contribution associated with uncorrelated noise.
As a result, moderate levels of common-mode noise primarily limit the absolute sensitivity but do not suppress the metrological gain $\eta(k_L)$.
\begin{figure}[htb!]
\centering
\includegraphics[width=\columnwidth]{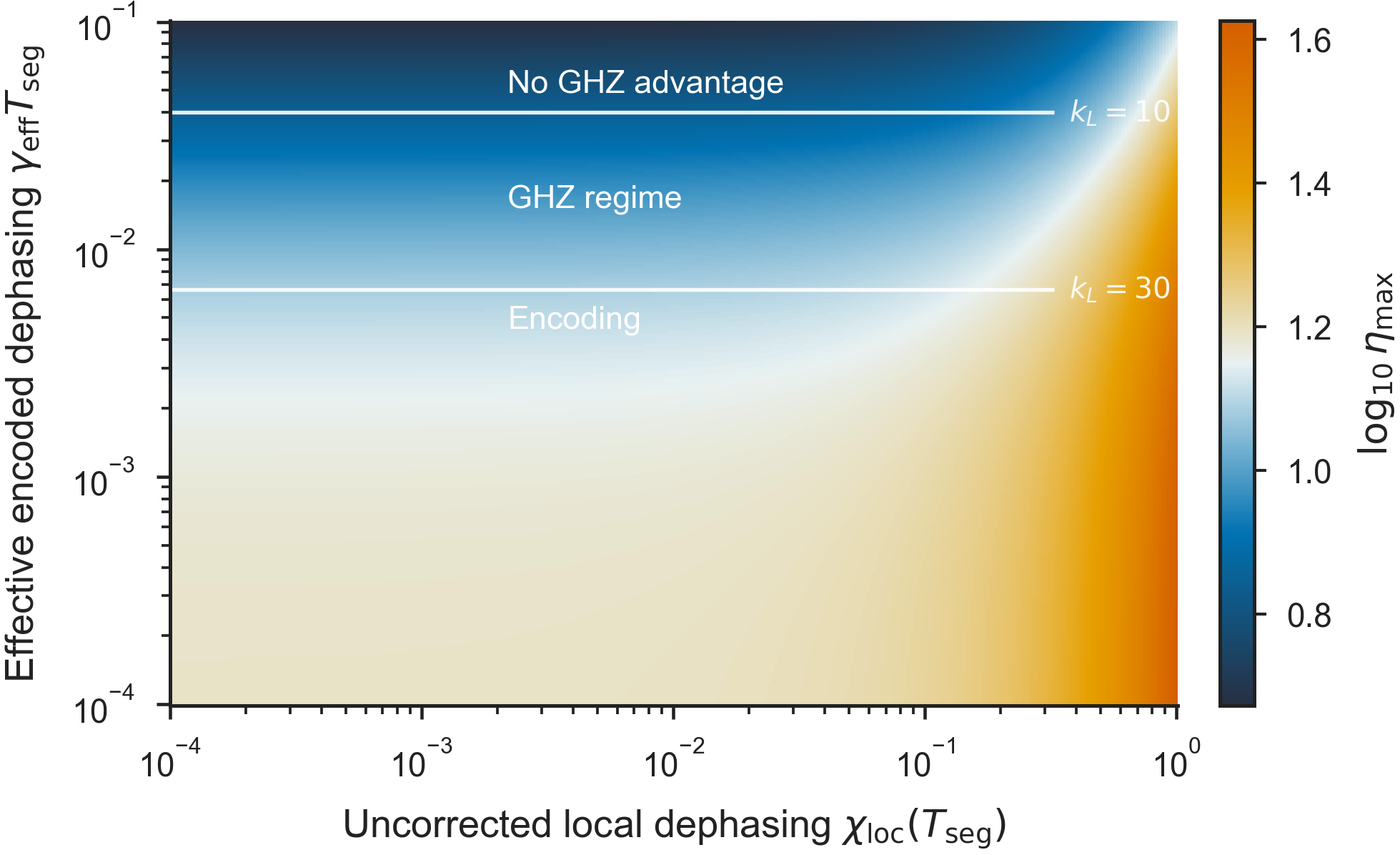}
\caption{Regimes of QEC benefit for entanglement-based axion sensing.
The color scale shows the maximal sensitivity enhancement factor $\eta_{\max}$ on a logarithmic scale as a function of the uncorrected single-qubit dephasing $\gamma_{\text{loc}}T_{\text{seg}}$ and the effective logical dephasing $\gamma_{\text{eff}}T_{\text{seg}}$.
White solid contours indicate the optimal logical GHZ size $k_L^*$, while the dashed contour marks $\eta_{\max}=1$.
The plot delineates distinct parameter regions corresponding to different balances between signal amplification and dephasing in QEC-protected entangled sensing.}
\label{fig:Regimes}
\end{figure}

Fig.~\ref{fig:Regimes} summarizes the regimes of QEC assisted entanglement-enhanced sensing by displaying both the optimal logical block size $k_L^*$ and the corresponding maximal enhancement factor $\eta_{\max}$ across representative device parameters.
In the absence of QEC, the accessible parameter space is confined mainly to the $k_L \approx 1$ region, indicating that entanglement provides little or no advantage due to rapid single-qubit dephasing.
Repetition-code QEC suppresses uncorrelated phase noise and shifts the system into a broad intermediate band with $k_L^* \sim 5$-$15$, consistent with the numerical optimization results presented in Sec.~\ref{sec:results}.
Within this window, multi-qubit GHZ states yield a robust enhancement over the SQL benchmark.
At larger values of $\gamma_{\text{eff}}T_{\text{seg}}$, the achievable block size is limited by effective logical dephasing; in this regime, large values of $\eta_{\max}$ reflect the strong suppression of the SQL reference rather than a high absolute sensitivity.

\subsection{Projected sensitivity improvement}
\label{subsec:results_sensitivity}

To connect directly to existing proposals, we apply the gain factor $\eta(m_a)$ to a baseline axion-electron coupling sensitivity curve, $g_{ae,{\text{min}}}^{\text{(base)}}(m_a)$, obtained from an unencoded spin-lock or frequency-modulated protocol with product-state sensing.
We denote by $\eta(m_a)\equiv \eta_{\max}(n_{\text{rep}}; m_a)$ the mass-dependent enhancement factor obtained by optimizing $\eta(k_L; m_a)$ over the logical GHZ size $k_L$, the QEC-enhanced sensitivity is simply
\begin{equation}
g_{ae,{\text{min}}}^{\text{(QEC)}}(m_a)=\frac{g_{ae,{\text{min}}}^{\text{(base)}}(m_a)}{\eta(m_a)}.
\end{equation}
This prescription is independent of the microscopic details of the baseline scheme and can be applied to any semiconductor spin qubit axion search whose Fisher information can be cast in the form of Eq.~\eqref{Eq:FISQL}.

\begin{figure}[htbp!]
\centering
\includegraphics[width=\columnwidth]{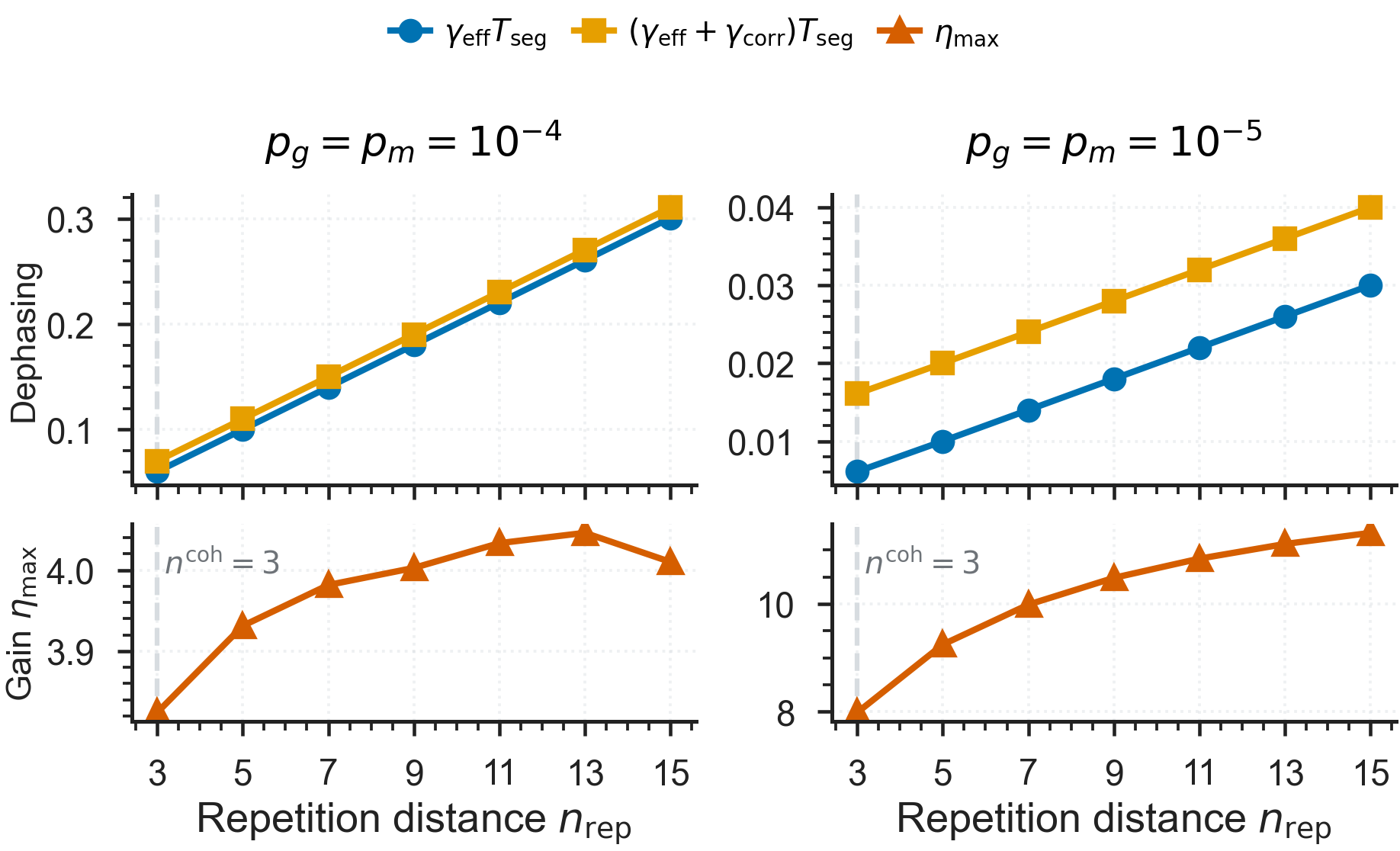}
\caption{Comparison of the logical dephasing and achievable metrological gain for realistic and optimistic control fidelities.
Left: $p_g=p_m=10^{-4}$ (future possible error parameters of Tab.~\ref{Tab: device_params}).
The logical dephasing $\gamma_{\text{eff}}T_{\text{seg}}$ is minimized at a small repetition distance $n_{\text{rep}}^{\text{coh}}\approx 3$, while the metrological gain $\eta_{\max}$ saturates at $\sim 5$ for $n_{\text{rep}}\gtrsim 10$ due to the measurement induced logical error floor.
Right: $p_g=p_m=10^{-5}$, representing a next-generation regime.
Both $\gamma_{\text{eff}}T_{\text{seg}}$ and $\eta_{\max}$ improve, with $\eta_{\max}$ reaching above~10, but the qualitative tradeoff and the location of the coherence-optimal code distance remain unchanged.}
\label{fig:gamma_eta}
\end{figure}

Fig.~\ref{fig:gamma_eta} summarizes how the balance between logical dephasing and metrological gain evolves with increasing repetition distance under realistic and improved control fidelities.
The baseline device parameters already place the system in a regime where uncorrelated phase noise can be strongly suppressed, but where the residual logical errors are governed primarily by measurement imperfections.
Consequently, enlarging the code yields only a modest increase in the usable GHZ block size and quickly enters a regime of diminishing returns.
Improved control fidelity shifts these trends quantitatively but does not alter the underlying structure of the trade-off, indicating that measurement fidelity defines the operational boundary for QEC-enhanced GHZ sensing and that moderate repetition distances are sufficient to achieve near-optimal performance.

The trends in Fig.~\ref{fig:gamma_eta} also indicate that repetition-code protection can suppress the intrinsic uncorrelated dephasing below the level set by control and readout faults.
Once this regime is reached, increasing the repetition distance yields diminishing returns because additional physical qubits introduce extra syndrome-extraction errors.
Correlated quasi-global noise then provides a separate lower bound that cannot be removed by the repetition code and would require a different mitigation strategy, such as differential sensing or decoherence-free subspaces.

Two-dimensional stabilizer codes, including the surface code, can in principle achieve far lower logical error rates.
 However, their application to sensing is constrained by architectural overhead: a large fraction of physical qubits serve as ancilla for stabilizer extraction and do not preserve the signal-bearing phase, reducing the number of qubits that contribute coherently to the GHZ state under a fixed hardware budget.
Furthermore, once the logical dephasing is limited by the correlated noise floor rather than by physical $Z$ errors, improvements to the logical error rate translate only into constant-factor changes in the Fisher information.
In contrast, the repetition-code approach preserves nearly all physical spins as active sensors while selectively mitigating the dominant noise channel, making it a resource-efficient choice for dephasing-limited
semiconductor platforms.
\begin{figure}[htbp!]
\centering
\includegraphics[width=\columnwidth]{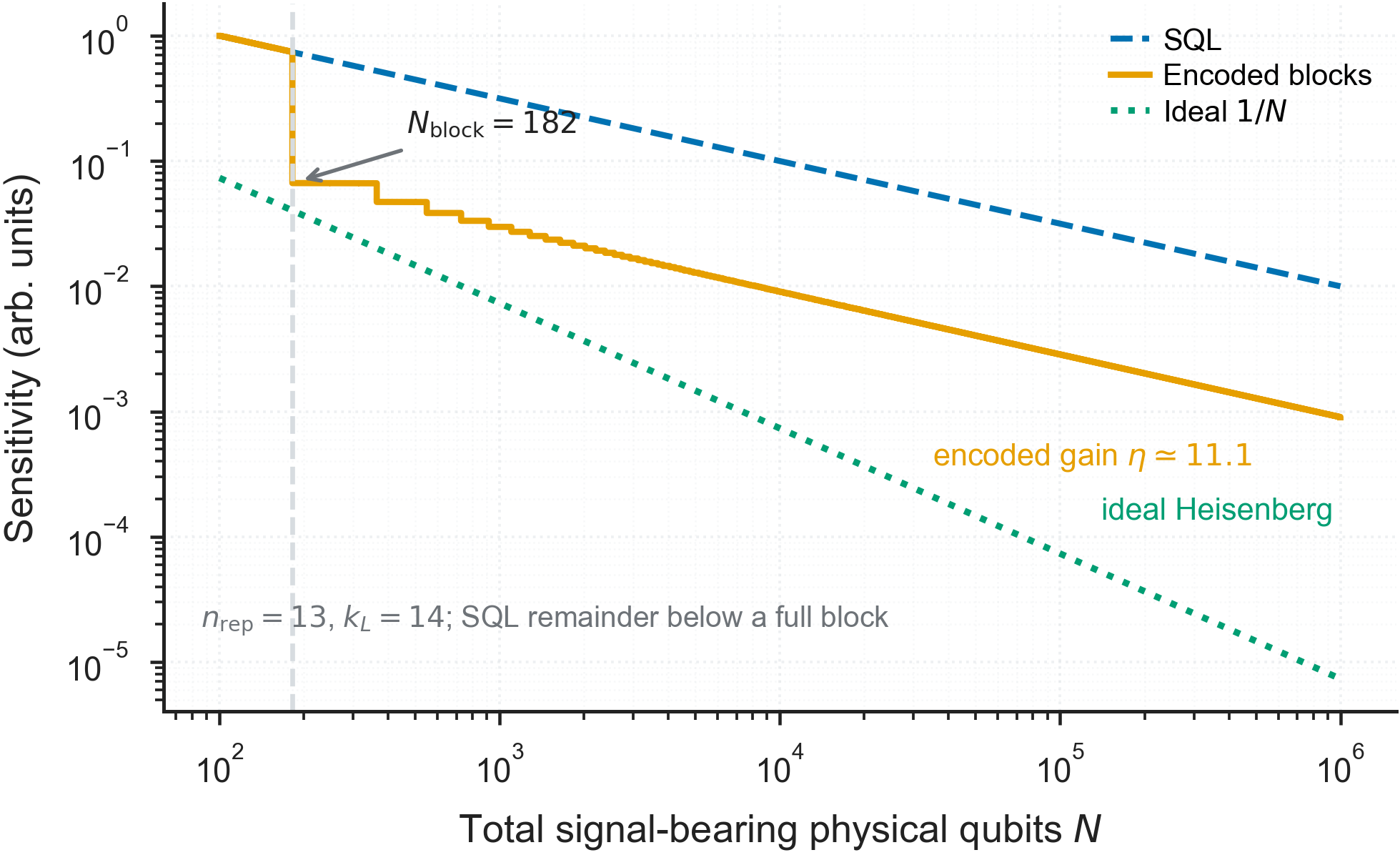}
\caption{Scaling of the projected axion-coupling sensitivity with the total number of signal-bearing physical qubits $N$.
The SQL benchmark follows the expected $N^{-1/2}$ behavior.
The encoded curve uses complete QEC-protected logical GHZ blocks with $(n_{\text{rep}},k_L)=(13,14)$, so the first complete encoded block appears at $N_{\text{block}}=n_{\text{rep}}k_L=182$.
For $N<N_{\text{block}}$, the encoded architecture has no complete block and therefore coincides with the SQL baseline.
For larger $N$, complete encoded blocks are instantiated discretely, while leftover physical qubits are treated as SQL resources, producing the visible staircase.
At large $N$, the encoded sensitivity approaches the SQL scaling with a constant downward shift corresponding to $\eta_{\text{seg}}\simeq 11.1$ in the matched-duty limit $R_{\text{duty}}=1$.
The ideal $1/N$ curve is shown only as a decoherence-free Heisenberg reference.
Thus QEC restores a finite entanglement advantage under realistic dephasing, but does not change the asymptotic qubit-number exponent.}
\label{fig:Sensitivity_VS_N}
\end{figure}

To determine the impact on a full-scale axion sensor, we tile the optimized logical GHZ blocks to construct an $N$-qubit device and evaluate the resulting minimum detectable coupling $g_{\min}(N)$.
The resulting scaling is shown in Fig.~\ref{fig:Sensitivity_VS_N}.
Both the SQL benchmark and the QEC-enhanced architecture retain the standard $N^{-1/2}$ dependence after block-size optimization.

The intuitive reason is that QEC restores a finite optimal entangled block size, rather than making the entire device behave as one decoherence-free $N$-qubit GHZ state.
For fixed device parameters, the optimal logical GHZ size $k_L^*$ and the repetition-code distance are set by the balance between coherent signal amplification, logical dephasing, residual correlated dephasing, and QEC-induced control faults.
Increasing the total number of physical qubits therefore increases the number of independent optimized blocks, rather than the size of a single coherent block.
Equivalently, if $N_{\text{block}}=n_{\text{rep}}k_L^*$ denotes the physical-qubit cost of one optimized encoded block, the large-$N$ Fisher information scales schematically as
\begin{equation}\label{Eq:block_fisher_scaling}
F_{\text{tot}}(N) \simeq \left\lfloor \frac{N}{N_{\text{block}}} \right\rfloor F_{\text{block}} + F_{\text{SQL}}^{\text{rem}},
\end{equation}
where $F_{\text{SQL}}^{\text{rem}}$ denotes the contribution from any leftover qubits not forming a complete encoded block.
Since $N_{\text{block}}$ and $F_{\text{block}}$ are fixed once the code and device parameters are optimized, $F_{\text{tot}}\propto N$ and therefore
\begin{equation}
g_{\min}(N)\propto F_{\text{tot}}^{-1/2}\propto N^{-1/2}.
\end{equation}
An intermediate scaling between $N^{-1/2}$ and $N^{-1}$ would require the optimal coherent block size itself to grow with $N$, which is prevented here by residual correlated dephasing and QEC-induced control overhead.

The ideal $1/N$ curve in Fig.~\ref{fig:Sensitivity_VS_N} is therefore shown only as a decoherence-free Heisenberg reference.
Under realistic device noise, the QEC-enhanced protocol provides a constant-factor downward shift relative to the SQL benchmark, not true Heisenberg scaling with qubit number.

\begin{figure}[htbp!]
\centering
\includegraphics[width=\columnwidth]{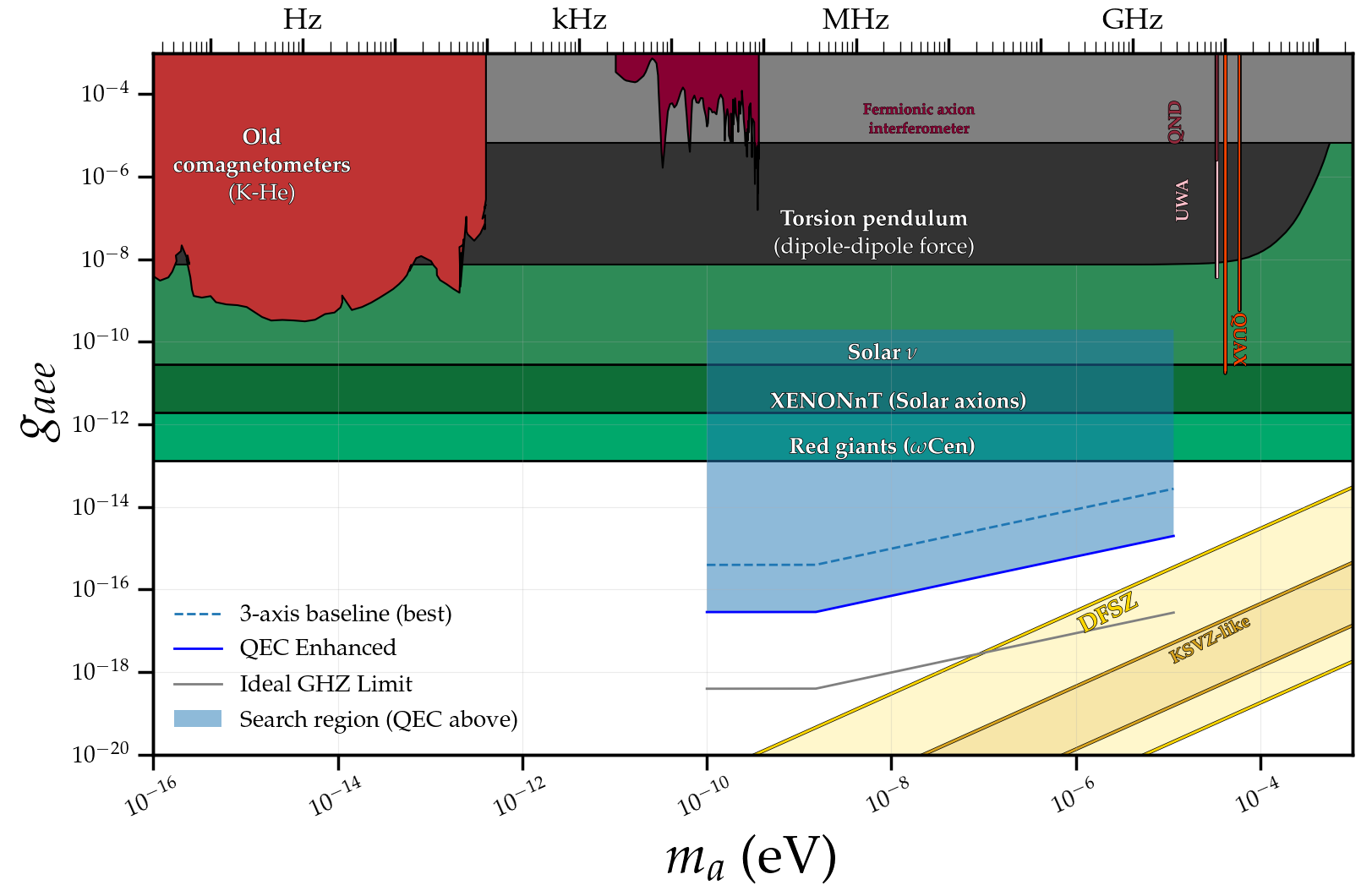}
\caption{Comparison of the SQL baseline axion–electron sensitivity $g_{ae,\text{min}}^{(\text{base})}(m_a)$ with the QEC-enhanced sensitivity $g_{ae,\text{min}}^{(\text{QEC})}(m_a)$, evaluated using the mass dependent enhancement factor $\eta(m_a)\sim 11.1$.
For $T_{\text{seg}}$ limited by hardware, the improvement of $\eta(m_a)$ matches Fig.~\ref{fig:optimal_k}, while at higher masses the reduced coherence time increases $\eta(m_a)$, widening the QEC advantage.
The SQL and ideal GHZ sensitivity baseline followed the results in Ref.~\cite{Tan2025IEEE, Tan2025PRD}.}
\label{fig:QEC_sensitivty}
\end{figure}

Fig.~\ref{fig:QEC_sensitivty} compares the baseline sensitivity $g_{ae,{\text{min}}}^{\text{(base)}}(m_a)$ to the QEC-enhanced sensitivity $g_{ae,{\text{min}}}^{\text{(QEC)}}(m_a)$ for the parameter set above.
In the mass range where $T_{\text{seg}}(m_a)$ is hardware-limited, the projected sensitivity curve is shifted downward by an approximately constant factor, corresponding to $\eta\simeq 10$.
This represents nearly an order of magnitude improvement in $g_{ae}$ sensitivity in the optimized high-fidelity regime, assuming identical physical hardware resources.
At higher masses, where $\eta(m_a)$ increases due to coherence time limited segmentation, the QEC-enhanced curve departs more strongly from the baseline and demonstrates that selective dephasing suppression can become even more valuable when the axion bandwidth is large.

\section{Conclusions}

We have shown that quantum error correction can substantially enhance axion search sensitivity in semiconductor spin qubit platforms by mitigating the dominant longitudinal dephasing that otherwise suppresses entanglement-enhanced sensing.
By integrating repetition-code protection with logical GHZ block entanglement, this approach suppresses the leading noise channel under the assumption that the coherent axion response is mapped to a code-preserving logical generator, yielding a hardware-efficient enhancement of the quantum Fisher information.
Under realistic semiconductor device parameters, modest code distances are sufficient to recover most of the achievable metrological gain.
More general two-dimensional codes can in principle achieve lower logical error rates, but at the cost of substantial overhead that is inefficient for sensing applications.
In dephasing-dominated, fixed-resource spin qubit architectures, repetition codes therefore provide an effectively optimal balance between noise suppression and signal accumulation \cite{Fowler2012, Riggelen2022, Mauricio2025}.
The framework developed here establishes a practical and broadly applicable pathway for incorporating quantum error correction into solid-state axion searches, and more generally into precision measurements based on semiconductor qubit platforms.
\clearpage

\end{document}